\documentclass[superscriptaddress, reprint, amsmath, amssymb, aps]{revtex4-1}
\pdfoutput=1
\usepackage{graphicx}
\usepackage{dcolumn}
\usepackage[english]{babel}
\renewcommand{\selectlanguage}[1]{}
\usepackage{bm}
\usepackage{soul}
\usepackage[a4paper,width=150mm,top=25mm,bottom=25mm,bindingoffset=0mm]{geometry}
\usepackage[dvipsnames]{xcolor}

\usepackage{chngcntr}
\usepackage{setspace}

\begin{document}

\title{
Mode visualisation and control of complex lasers using neural networks}

\author{Wai Kit Ng}
\affiliation{Blackett Laboratory, Department of Physics, Imperial College London, London, UK}
\author{T. V. Raziman}
\affiliation{Blackett Laboratory, Department of Physics, Imperial College London, London, UK}
\affiliation{Department of Mathematics, Imperial College London, London, UK}
\author{Dhruv Saxena}
\affiliation{Blackett Laboratory, Department of Physics, Imperial College London, London, UK}
\author{Korneel Molkens}
\affiliation{Physics and Chemistry of Nanostructures (PCN), Ghent University, Krijgslaan 281-S3, B9000 Gent, Belgium}
\affiliation{Center for Nano- and Biophotonics, Ghent University, 9052 Ghent, Belgium}
\affiliation{Photonics Research Group, Ghent University - imec, Technologiepark-Zwijnaarde 126, 9052 Ghent, Belgium}
\author{Ivo Tanghe}
\affiliation{Physics and Chemistry of Nanostructures (PCN), Ghent University, Krijgslaan 281-S3, B9000 Gent, Belgium}
\affiliation{Center for Nano- and Biophotonics, Ghent University, 9052 Ghent, Belgium}
\affiliation{Photonics Research Group, Ghent University - imec, Technologiepark-Zwijnaarde 126, 9052 Ghent, Belgium}
\author{Zhenghe Xuan}
\affiliation{Blackett Laboratory, Department of Physics, Imperial College London, London, UK}
\author{Pieter Geiregat}
\affiliation{Physics and Chemistry of Nanostructures (PCN), Ghent University, Krijgslaan 281-S3, B9000 Gent, Belgium}
\affiliation{Center for Nano- and Biophotonics, Ghent University, 9052 Ghent, Belgium}
\author{Dries Van Thourhout}
\affiliation{Center for Nano- and Biophotonics, Ghent University, 9052 Ghent, Belgium}
\affiliation{Photonics Research Group, Ghent University - imec, Technologiepark-Zwijnaarde 126, 9052 Ghent, Belgium}
\author{Mauricio Barahona}
\affiliation{Department of Mathematics, Imperial College London, London, UK}
\author{Riccardo Sapienza}
\affiliation{Blackett Laboratory, Department of Physics, Imperial College London, London, UK}
\email{r.sapienza@imperial.ac.uk}

\begin{abstract}
Understanding the behaviour of complex laser systems is an outstanding challenge, especially in the presence of nonlinear interactions between modes. Hidden features, such as the gain distributions and spatial localisation of lasing modes, often cannot be revealed experimentally, yet they are crucial to determining the laser action. Here, we introduce a lasing spectroscopy method that can visualise the gain profiles of the modes in complex lasers using an artificial neural network. The spatial gain distributions of different lasing modes in a disorderly coupled microring array are reconstructed without prior knowledge of the laser topology. We further extend the neural network to a tandem neural network that can control the laser emission by matching the modal gain/loss profile to selectively enhance the targeted modes. This mode visualisation method offers a new approach to extracting hidden spatial mode features from photonic structures, which could improve our understanding and control of complex photonic systems.
\end{abstract}

\maketitle

\section*{Introduction}

Complex photonic structures are feature-rich both in their material architecture and optical response. Complex laser systems, such as random lasers \cite{caoRandomLaserAction1999, caoUltravioletLasingResonators1998, trivediSelforganizedLasersReconfigurable2022, turitsynRandomDistributedFeedback2014, bachelardAdaptivePumpingSpectral2014}, wave-chaotic cavities \cite{bittnerSuppressingSpatiotemporalLasing2018, reddingLowSpatialCoherence2015}, and coupled lasers \cite{hodaeiParitytimeSymmetricMicroring2014a, qiaoHigherdimensionalSupersymmetricMicrolaser2021, zhaoTopologicalHybridSilicon2018}, exhibit properties including broad spectrum and low coherence. These properties cannot be easily replaced by conventional lasers \cite{caoComplexLasersControllable2019} and are favoured in applications such as sensing \cite{wanismailDopamineSensingMeasurement2016, caixeiroSilkBasedBiocompatible2016} and speckle-free imaging \cite{reddingSpecklefreeLaserImaging2012, farrokhiHighbrightnessLaserImaging2017}. 
However, the complexity and nonlinearity of these structures make their emission spectra and profiles challenging to predict and control. 

Coupled nano- and micro-laser systems bring together simple resonator units to form rich modes with non-trivial laser dynamics \cite{bandresTopologicalInsulatorLaser2018, hamelSpontaneousMirrorsymmetryBreaking2015, kodigalaLasingActionPhotonic2017, qiaoHigherdimensionalSupersymmetricMicrolaser2021, qiaoPhotoisomerizationActivatedIntramolecular2020, teimourpourNonHermitianEngineeringSingle2016, zhangPhotonicSkinsBased2021, zhaoTopologicalHybridSilicon2018}. Yet, they are challenging to understand and control using existing methods. Simulating the optical modes of coupled lasers would be computationally infeasible unless the design is simplified with strong symmetries and weak interactions \cite{hodaeiParitytimeSymmetricMicroring2014a, benzaouiaNonlinearExceptionalpointLasing2022}, but this would inevitably reduce the functionalities of the lasers. Experimentally estimating the mode profiles through far-field emission images would also be limited to cases of unidirectional emission or simple lasing modes~\cite{lafalceRobustLasingModes2019}. In some special cases, such as by adding additional cavities to a coupled laser \cite{qiaoHigherdimensionalSupersymmetricMicrolaser2021} or by unbalancing system excitations \cite{hodaeiParitytimeSymmetricMicroring2014a, fischerControllingLasingExceptional2024,saxenaSensitivitySpectralControl2022}, it is possible to control lasers to output a single-mode emission with limited tunable range. However, performing lasing control over a wide frequency range on disordered and symmetry-broken structures is still challenging. All these limitations call for new methods to study and control such complex laser systems. 

Over the last three decades, machine learning (ML) \cite{jordanMachineLearningTrends2015}, has demonstrated the potential to provide new insight into complex systems. Recently, ML has become a crucial tool for physics as it can provide inverse design capabilities without requiring a complete physical model. In photonics, ML has been used extensively in the optimisation of complex systems, especially for device design \cite{chughMachineLearningRegression2019, liuTrainingDeepNeural2018, maDeepLearningEnabledOnDemandDesign2018, malkielPlasmonicNanostructureDesign2018, pilozziMachineLearningInverse2018, sajedianOptimisationColourGeneration2019, soOndemandDesignSpectrally2021, tahersimaDeepNeuralNetwork2019, turduevUltracompactPhotonicStructure2018}. 
Going beyond ``black box" operation, there is a growing interest in explainable artificial intelligence (XAI) \cite{adadiPeekingBlackBoxSurvey2018}. Such approaches allow us to gain model transparency, helping us to understand how ML interprets the problems at hand and, thus, learn from the ML model. For example, the visual explanation of computer vision \cite{selvarajuGradCAMVisualExplanations2020} and feature visualisation of CNN-based Raman spectrum analysis \cite{fukuharaFeatureVisualizationRaman2019} have shown how and what the important regions are in their corresponding recognition tasks, providing us with valuable insight into formulating physical models. By applying XAI to complex laser studies, the complex processes and features that are currently experimentally inaccessible could be revealed by unfolding an ML model in a similar manner.

\begin{figure*}[!htb]
    \centering
    \includegraphics[width= 1\textwidth]{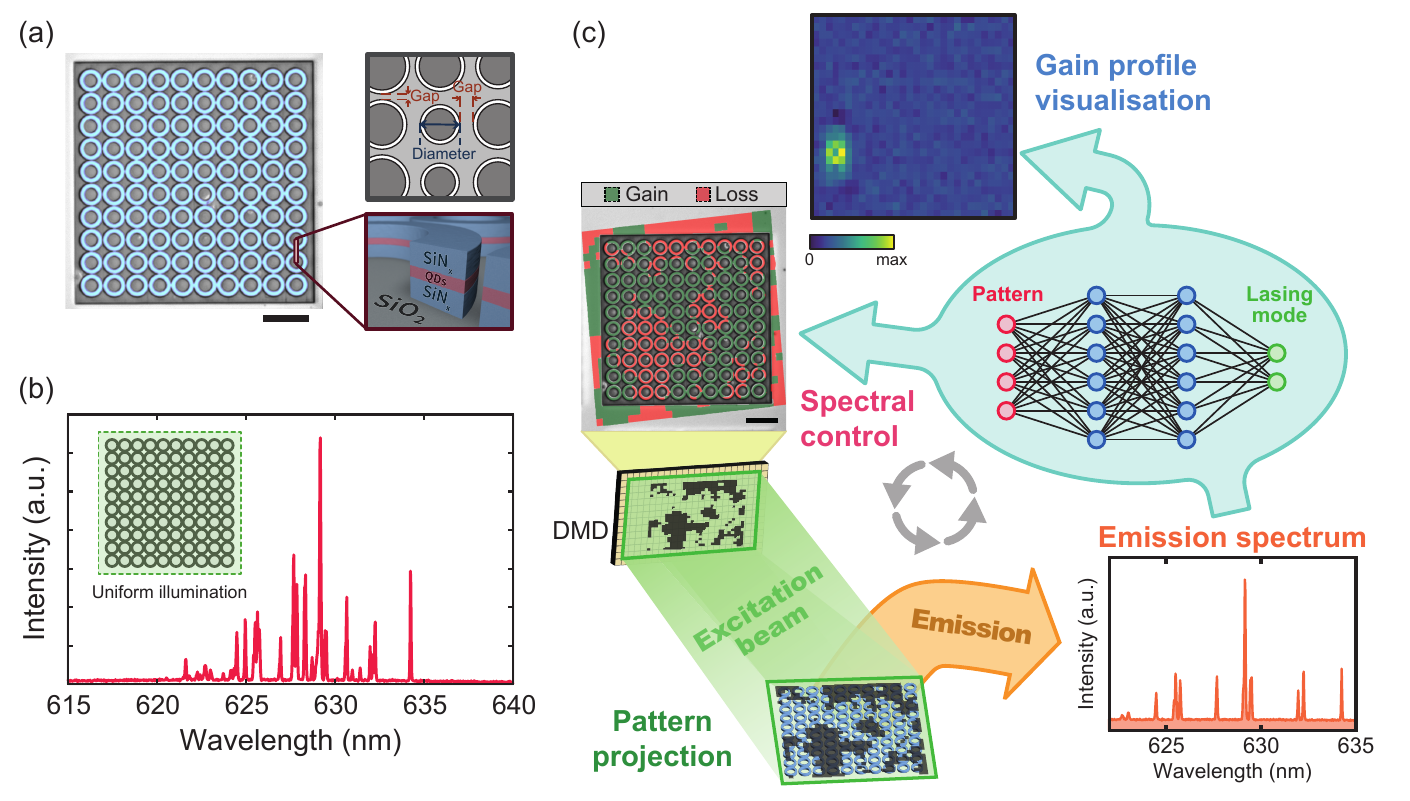}
    \caption{\textbf{Uncovering complex lasing modes with neural networks. }
    (a) The pseudo-colour bright field image (left) and schematic sketch (top right) of a disorderly coupled 10$\times$10 QD-SiN microring array with varying gaps and diameters. For detailed characteristics of the structure and material, see Supplementary FIG. \ref{SUPPFIGURE_RingChars}. The cross-section of a microring (SiN/QD/SiN stack on $\rm SiO_{2}$ substrate) is illustrated in the bottom right.
    (b) The lasing spectrum of the disorderly coupled laser array under uniform excitation, which shows a complex and multi-mode lasing behaviour. 
    (c) Gain profile visualisation and spectral control through artificial neural networks. The excitation beam, patterned via a digital micromirror device (DMD), illuminates the microring array, resulting in different emission spectra. The modal spatial gain profile can be visualised from the emission spectrum by unfolding the neural network model. To control laser emission, the excitation pattern of a spectrum can be predicted by solving the inverse problem through a tandem neural network. Scale bars are 20~$\mu$m.}
    \label{FIGURE01}
\end{figure*}

Here we use ML to help unveil the spatial and spectral features of a complex laser system.  In particular, we train and unfold an ML network model to reveal the origins of the lasing modes in a disorderly coupled microring array and control its lasing spectra. The spatial gain profiles of the lasing modes are identified by reverse engineering the connection weights in a multi-layer perceptron (MLP) model. Having produced a picture of the individual gain distributions of the modes, we then extend the MLP into a tandem neural network (TNN) that enables spectral control in single-mode and dual-mode emissions by effectively suppressing unwanted modes. Our results in mode selection and mode profile visualisation extend the use of ML in photonic systems beyond design optimisation, towards the understanding and control of photonic devices. 

\section*{Results}

Silicon nitride (SiN) microdisks with quantum dots (QDs) have previously been shown to be multi-mode resonators with a high Q factor and vast design flexibility \cite{xieChipIntegratedQuantum2017}. On the same hybrid QD-SiN platform, we fabricated a microring array with $10 \times 10$ microrings as a disorderly coupled laser system \cite{xieChipIntegratedQuantum2017, zhuOnChipSingleModeDistributed2017} (FIG.~\ref{FIGURE01}(a), see Methods). To realise a weakly coupled complex laser system, the microrings were designed to have slightly different diameters and ring-to-ring gaps to achieve non-uniform resonance frequencies and coupling strengths (see \ref{Sample_char}). In cases where the adjacent microrings have similar resonance frequencies (diameters) at a close distance, the microrings are coupled evanescently. The resulting collective modes in the system involve contributions from several rings. Hence, the supported modes have similar thresholds, and they overlap and compete for gain in a complex nonlinear system. Compared to a single microring under uniform illumination (FIG.~\ref{SUPPFIGURE_RingChars}(b)), the mode competition in the microring array results in a far more complicated lasing spectrum with more than 25 distinct lasing peaks within a 10 nm spectral range (FIG.~\ref{FIGURE01}(b)).

In the disorderly coupled microring array, the modes extract gain from different areas and exhibit distinct spatial distributions due to mode competition. Therefore, the lasing modes can be manipulated by modifying the gain profile of the system through spatially patterned excitations \cite{qiaoAdaptivePumpingSpectral2018, bachelardAdaptivePumpingSpectral2014, leonettiActiveSubnanometerSpectral2013, liewPumpcontrolledModalInteractions2015, saxenaSensitivitySpectralControl2022}. By selectively illuminating the array with different patterns via a programmable digital micromirror device (DMD), the excited QDs embedded in the selected regions provide gain, while the remaining regions are lossy due to QD absorption (see Methods). This patterned input allows for the modulation of the internal gain distribution to selectively enhance and suppress modes through mode competition in a system with a fixed topology. In such a disordered system, obtaining the gain distributions of the modes is a complex problem, so we established a learnable model via a neural network.

To visualise and control the lasing modes, neural network models were built to connect the gain distributions of the modes to the lasing spectra (Fig. \ref{FIGURE01}(c)). The neural networks, trained with both spatial gain patterns and emission spectra, can be unfolded to extract the modal gain profiles. With the gain profiles of each mode (i.e., each peak in the spectrum), the complex illumination pattern needed to control a particular spectrum (with a combination of modes) can be generated by solving this inverse problem with a two-step tandem neural network. This can be achieved without extra effort to acquire additional experimental training data.

The central idea behind visualising the gain profile of the modes is to reveal the modes' spatial information as encoded in a trained nonlinear MLP model. The contributions of spatial pixels to a mode can then be traced back via the inner connections between layers. For simplicity, consider first a single-layer neural network model, which can be written in matrix form as
\begin{equation}\label{Single_NN}
\mathbf{M} = \sigma(\mathbf{W} \cdot \mathbf{P} + \mathbf{B}), 
\end{equation}
where $\mathbf{M} = [m_{1}, ..., m_{M}]$ is the vector of one-hot encodings for $M$ modes ($m_{i} \in \{0,1\}$),
$\mathbf{P} = [p_{1}, ..., p_{I}]$ is the vector containing the intensities of the $I$ spatial pixels ($p_{i} \in \mathbb{R}$), $\mathbf{B}$ is a bias vector, and $\sigma(\cdot)$ is a nonlinear activation function (e.g., sigmoid or ReLU). The nonlinearity of the activation function is the key to building a powerful and accurate model that maps the input (pixel intensities) to the output (the lasing state of a mode). Yet, for the same reason, the activation function also introduces non-invertibility so that the spatial gain profile cannot be directly obtained from the spectrum. 

In our system, however, the gain profile for each lasing mode can be estimated by considering how influential each spatial pixel is just below the lasing threshold of the mode. For a pixel that overlaps well with the mode, a small change in pixel intensity can significantly alter the state of the mode, taking it above the lasing threshold. In contrast, if the pixel does not overlap with the mode (i.e., it does not contribute to the gain), whether the mode lases or not is irrelevant to that pixel's intensity.

To see how a selected mode $m_{i}$ changes with respect to changes in the intensity of pixel $p_{j}$, it follows from Equation \ref{Single_NN} that 
\begin{equation}\label{derivative_single_NN}
\frac{\partial m_{i}}{\partial p_{j}} = \frac{\partial \sigma \left ( \widetilde{m}_{i} \right )}{\partial \widetilde{m}_{i}} \, \mathbf{W_{ij}},
\end{equation}
where $\widetilde{m}_{i} = $ $\sum_{j} \mathbf{W_{ij}}~p_{j} + b_{i}$ and $\mathbf{W_{ij}}$ is the weight matrix element. For a multi-label, single-class classification neural network model, which is the type of model used in this work, the partial derivative $\frac{\partial \sigma \left ( \widetilde{m}_{i} \right )}{\partial \widetilde{m}_{i}}$ is then independent of $p_j$ and only determined by how close the system is to the threshold of mode $m_{i}$. As the system populates closer to the threshold, the contribution from pixel $p_{j}$ becomes more significant due to the increase in the partial derivative, i.e., the system becomes more sensitive. Therefore, for the target mode $m_{i}$, the partial derivative term is constant across each pixel $p_{j}$, but the weight matrix element $\mathbf{W_{ij}}$ differs among the pixels. As a result, the weight matrix element $\mathbf{W_{ij}}$ becomes the only important factor left in distinguishing how influential each spatial pixel is compared to the others, thus visualising the gain profile of mode $m_{i}$.

This relationship extends to neural networks with more than one layer (see \ref{MultiLayer_theory}) provided that the neural network has local activation functions (e.g., sigmoid or ReLU). Yet, for deeper networks with a large number of layers, the assumption that all relevant intermediate activation functions in the hidden layers activate simultaneously might not be satisfied, thereby increasing inaccuracies in determining the gain profile. From experience, our approach works best for relatively shallow MLPs with 3 or fewer layers.

\begin{figure*}[!htb]
    \centering
    \includegraphics[width= 1\textwidth]{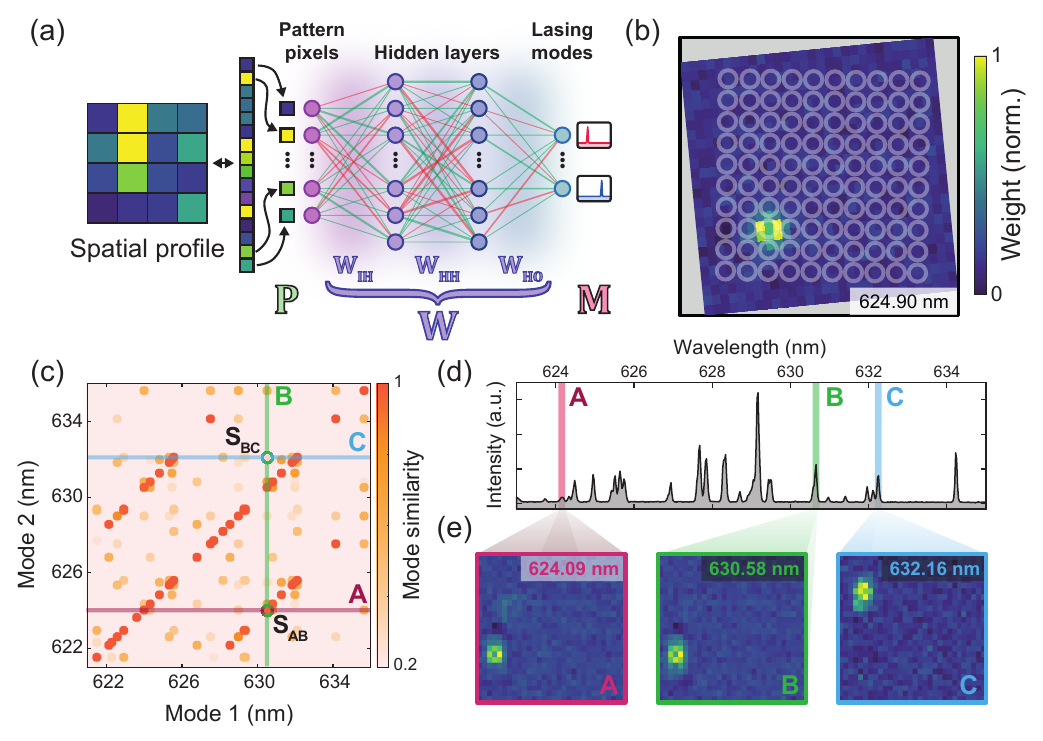}
    \caption{\textbf{Mode visualisation using a multi-layer perceptron.}
    (a) The mode visualisation scheme via multi-layer perceptron (MLP) neural network. The network connects the excitation patterns to the emission modes. The spatial gain profile $\mathbf{P}$ of each mode can be estimated from the network weight matrices ($\mathbf{W} = \mathbf{W_{HO}} \cdot \mathbf{W_{HH}} \cdot \mathbf{W_{IH}}$).
    (b) A visualised gain profile for a lasing mode at 624.90 nm with the microring sample structures superimposed.
    (c) The spatial similarity map between all the gain profiles of the lasing modes. The highly similar mode pairs form super- and sub-diagonal lines representing a constant spectral distance between the modes. Two pairs of modes (AB and BC) with high and low similarities are also highlighted.
    (d) Three spectrally uncorrelated modes (A, B, and C) are highlighted in the lasing spectrum of the disorderly coupled microring array. Their spatial mode profiles are displayed in (e). The similarity between the modes can be found by crossing two corresponding lines in (c). This shows a high similarity between modes A and B ($\rm S_{AB} = 0.9577$), but not between modes B and C ($\rm S_{BC} = -0.0473$).}
\label{FIGURE02}
\end{figure*}

To visualise the spatial gain profile of the disorderly coupled array, we trained a spectral prediction network (SN), an MLP with 2 hidden layers (1024 and 64 nodes respectively), shown in FIG.~\ref{FIGURE02}(a). To train the network, we collected the lasing responses from the array with a set of random Perlin noise illuminations \cite{perlinImageSynthesizer1985}. The Perlin noise patterns are projected through a DMD and cover the entire area of the array with some degree of clustering (see FIG.~\ref{FIGURE01}(c)) to sample across different gain-loss distributions. The experimentally collected excitation patterns and lasing spectra are then used as matched pairs to train the SN offline (see \ref{Data_preprocess} and \ref{SN_training}). 

From the trained network model, we obtain the overall weight matrix ($\mathbf{W} = \mathbf{W_{HO}} \cdot \mathbf{W_{HH}} \cdot \mathbf{W_{IH}}$) that allows us to picture the spatial gain profiles of different lasing modes, as illustrated in an example of the reconstructed mode at 624.90 nm (FIG.~\ref{FIGURE02}(b)). Note how the recovered gain profile shows a localised circular structure that matches a microring resonator in the array. This also matches our expectations of the mode distributions from the disorderly coupled microring laser. Due to the weak coupling in the highly disordered microring array, the modes with the lowest threshold are likely to be those that are more localised in a single ring. This example demonstrates how using a simple neural network model allows us to visualise the underlying spatial profile of lasing modes.

Using this ML approach provides us with a new perspective to see the spatial modal gain profiles of laser systems. FIG.~\ref{FIGURE02}(c) shows the spatial similarities of all the lasing modes in the same microring array. Many pairs of modes in the array show strong spatial overlap with a constant spectral distance between them (i.e., the super- and sub-diagonal bright orange points in FIG.~\ref{FIGURE02}(c)). To illustrate the relationship between mode pairs, 2 pairs of modes (AB and BC) are highlighted with their recovered spatial profile (mode A, B, and C) in FIG.~\ref{FIGURE02}(e). From the complex emission spectrum of the array (FIG.~\ref{FIGURE02}(d)), it is hard to tell if these modes are competing. However, in their spatial maps, mode A (624.09 nm, magenta) is nearly identical to, and colocalised with, mode B (630.58 nm, green) with a cosine similarity of 0.9577, but not with mode C (632.16 nm, blue). This indicates that modes A and B share the gain in the same microring and are therefore likely to be spatially co-localised. Indeed, the spectral separation between the modes is measured as 6.49 nm (FIG.~\ref{FIGURE02}(d)), which is close to the free-spectral range of the whispering gallery mode (WGM) of a 10~$\mu$m microring resonator. Mode A is therefore very likely the higher-order mode of mode B, which is mainly located in the same resonator. This result also implies that, despite being coupled, the WGMs in our microring array are only weakly perturbed because of the large detuning of resonance frequencies as well as weak local coupling strengths due to structural disorder.

\begin{figure*}[!htb]
    \centering
    \includegraphics[width= 1\textwidth]{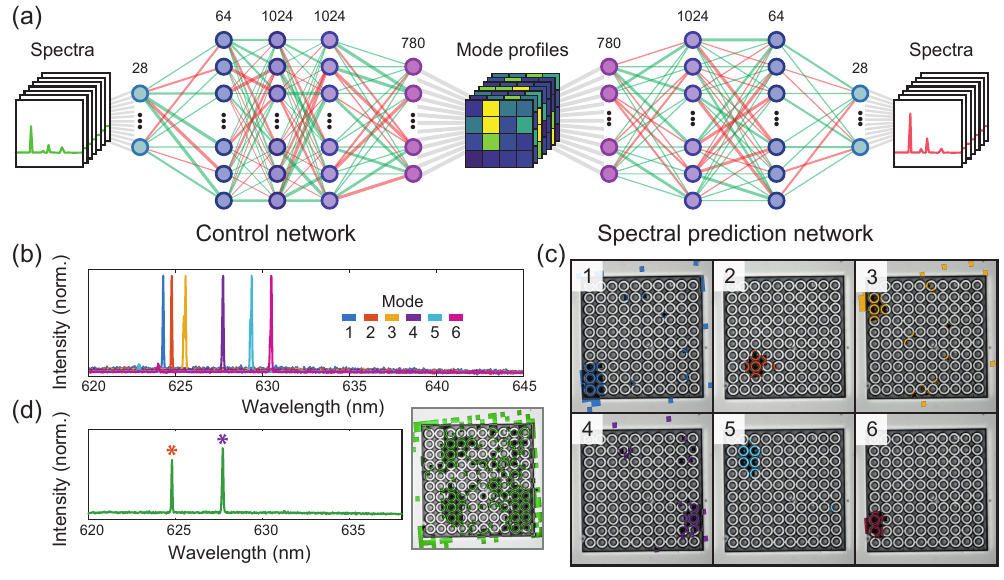}
    \caption{\textbf{Lasing control using a tandem neural network (TNN).}
    (a) The architecture of the TNN, which combines 2 artificial neural networks --- the control network and the spectral prediction network --- for lasing control. The number of nodes for each layer is shown at the top. The control network stands alone as a model to predict the excitation pattern required for the targeted modes after training.
    (b) Controlled single-mode emissions on the disorderly coupled microring array, with the best side-mode suppression ratio (SMSR) of 14.41 dB. 
    (c) The predicted excitation profiles for different single-mode emissions in (b). The target lasing modes of the spectra are (1) 624.41 nm, (2) 624.90 nm, (3) 625.67 nm, (4) 627.78 nm, (5) 629.44 nm, and (6) 630.58 nm. 
    (d) Dual-mode (624.90 nm and 627.78 nm) lasing control performed on the same system using the same model. The asterisks indicate the spectral positions of the target modes. The right panel shows the corresponding predicted illumination pattern, which is complex and not equivalent to the sum of two corresponding single-mode profiles (modes 2 and 4 in (c)).}\label{FIGURE03}
\end{figure*}

Beyond the mode visualisation task described above, the spatial information captured by the neural network can also be applied to the task of controlling complex lasers. Importantly, although illuminating a system with the recovered spatial profile allows us to excite the mode by providing maximum gain, it does not guarantee the suppression of other modes --- undesired modes that have a high degree of spatial overlap with the target mode might still emerge. Furthermore, the overall mode profiles for multi-mode lasing differ from the sum of individual single-mode profiles --- a nonlinear effect due to complex mode competition in the coupled system. Therefore, to enable the control of the laser system, it is necessary to develop a model that can predict the excitation profile directly from the lasing spectra with consideration of mode competition.

Finding an illumination profile that gives rise to a desired lasing spectrum is an inverse problem, which we solve here by extending the SN into a tandem neural network (TNN). The TNN is built on the original SN appending another MLP, the control network (CN), joined at the front (Fig.~\ref{FIGURE03}(a)). The CN consists of 3 hidden layers (64, 1024, and 1024 nodes respectively) where spectral modes are the inputs and illumination patterns are the output; hence, the CN acts as a predictor for the excitation profile from the desired modes based on the knowledge established in the SN. The SN and CN are then trained with the same set of experimental data in sequence (see \ref{TNN_training}). After training, the CN becomes a standalone model capable of controlling the lasing mode by generating the required excitation patterns. In FIG.~\ref{FIGURE03}(b), (c), we demonstrate the control of various single mode emissions from the microring array by exciting the array with the generated patterns. The single-mode emissions are achieved with the best side-mode suppression ratio (SMSR) of 14.41 dB at 628.20 nm. Note that, compared to the mode visualisation scheme with the SN (FIG. \ref{FIGURE02}(e)), the CN does not provide high-resolution spatial profiles that recover the fine gain structure of the modes. Instead, the CN focuses on enhancing the target mode while suppressing the side modes, thus resulting in slightly different, more extended excitation patterns.

Apart from single-mode emissions, multi-mode emission can also be controlled using the same inverse neural network model. We demonstrated dual-wavelength lasing control of the lasing emission at 624.9 nm and 627.7 nm from the same array using the same predictor with no extra training (see FIG.~\ref{FIGURE03}(d)). As a result of the competition among the modes
, the predicted excitation profile is complex and differs strongly from the linear sum of the two individual mode profiles (mode profiles 2 and 4 in FIG. \ref{FIGURE03}(c)). 

\section*{Discussion}

Compared to other spectroscopy methods, such as wide-field microscopy, which estimates the mode spatial profiles by imaging their emission, the mode visualisation method we propose collects no spatial information from the emission. It is therefore not affected by the emission directionality or the light scattering angles. This allows our method to be applied to any laser or ASE structure regardless of their outcoupling of emission. 

Establishing a lasing prediction model based on the mode visualisation model enables lasing control in multiple frequencies, a desirable feature for applications that require the simultaneous manipulation of lasing peaks, such as multiplexing communication and remote sensing by differential absorption LIDAR \cite{wagnerMultifrequencyDifferentialAbsorption2018}. Since the training and inference of the neural network can be done offline, this approach could also enable the fast switching of complex lasers between different sets of modes down to a few picoseconds (limited only by the carrier relaxation time). 

Although the methods described here could be applied to highly multimodal lasers, the efficiency of the lasing control is expected to drop with increasing complexity and number of modes. As with other ML problems, modelling the lasing systems with more modes would require more training data \cite{raudysSmallSampleSize1991}, which would usually result in either sacrificing the profile resolution or requiring a longer data acquisition time. Also, the ML approach relies on consistent and reproducible experimental datasets. Therefore, lasing materials that change over time, such as laser dyes that gradually degrade under illumination, would limit the prediction accuracy. Stable gain media, such as semiconductors and embedded colloidal QDs, would thus be more suitable for mode visualisation and lasing control with the proposed ML methods. To further improve the scheme, training the neural networks with simulation data, if a good physical model is available \cite{saxenaSensitivitySpectralControl2022}, could be an alternative strategy that could also speed up the training process and reduce device degradation. 

\section*{Conclusion}
In conclusion, we have shown how artificial neural networks can visualise the modal spatial gain profiles and be used to control the lasing modes from a disorderly coupled microring array. By training the neural network with random excitation patterns, the spatial gain profiles encoded in the network can be mapped by unfolding the connections between neural network layers. We also demonstrated lasing control using tandem neural networks, achieving a wide range of single-mode and dual-mode emissions from the same device without extra data collection. Although we focused on mode visualisation and control in weakly coupled lasers, these methods can also be applied to systems with more complex lasing modes, such as network lasers and random lasers. 
This work with neural networks opens the possibility of developing spectroscopic tools to understand the hidden features in complex laser systems, with potential applications in optical information processing, smart illumination, and optical computing. 

\section*{Methods}

\subsection*{Microring array fabrication}
The disorderly coupled microring array was fabricated as reported in \cite{xieChipIntegratedQuantum2017}.  A bottom layer of 130 nm thick SiN was first deposited on a $\mathrm{SiO_{2}}$ wafer through low radiofrequency plasma-enhanced chemical vapour deposition (PECVD). Then, a layer of 55 nm thick CdSe/CdS quantum dots (QDs) was spin-coated. The QDs were purified and dispersed in toluene, with adjusted concentration for a desired QD layer thickness. After that, a 105 nm thick SiN top layer was deposited on the QDs as a crack-free encapsulation layer via a mixed radiofrequency PECVD process. Subsequently, photolithography was employed to define the microring structures by patterning the resist on SiN as a mask. Finally, a specifically optimised reactive ion etching (RIE) process based on a $\rm CF_{4}/H_{2}$ gas mixture was used to etch the SiN/QD/SiN layers to attain microring arrays with smooth and steep sidewalls.

\subsection*{Optical experiment}
The microring arrays were gain-modulated and characterised in an optical microscopy setup via selective excitation. The excitation source, a Nd:YAG pulsed laser (TEEM Power-Chip, $\lambda$ = 532 nm, pulse width 400 ps, energy per pulse 20 $\mu$J), was patterned with a programmable digital micromirror device (DMD, Ajile AJD-4500) and projected onto the sample through a Nikon Ti microscope mounted with a 20$\times$ objective lens (Nikon CFI Super Fluor 20$\times$, 0.75 N.A., 1.0 mm WD). The laser beam was initially expanded by a 3$\times$ telescope before patterning to achieve uniform excitation fluence for the pixels. For any partial illumination patterns, the overall excitation fluence delivered on the sample was held constant (10.24~mJ cm$^{-2}$~pulse$^{-1}$), and the DMD micromirrors were grouped into superpixels with an area of (4 $\times$ 4.7)~$\mu$m$^{2}$ each, which is about one-quarter of the size of the microrings. The lasing emission from the microring array was then collected through the same objective lens, filtered, and further focused by a cylindrical lens into a vertical stripe on the spectrometer entrance slit to maximise the signal-to-noise ratio. The lasing spectrum was spectrally analysed with a grating spectrometer (Princeton Instruments Isoplane-320) equipped with 1800 grooves/mm holographic grating (0.05 nm resolution) and a charge-coupled device camera (CCD, Princeton Instruments Pixis 400).

\section*{Acknowledgements}
W. K. N. acknowledges the research support funded by the President’s PhD Scholarships from Imperial College London. K. M. acknowledges FWO-Vlaanderen for research funding (FWO project G0B2921N). The authors also acknowledge the support of EPSRC (EP/T027258/1).

\section*{Author contributions}
Conceptualization: WKN, TVR, DS, ZX; optical measurements: WKN, DS; theory: WKN, TVR, MB; data analysis: WKN, TVR; fabrication: KM, IT; visualization: WKN; funding acquisition \& project administration: PG, DVT, RS; supervision: RS; writing (original draft): WKN; writing (review and editing): WKN, TVR, DS, KM, IT, ZX, PG, DVT, MB, RS.

\section*{Competing interests}
The authors declare no competing interests.

\section*{Materials \& Correspondence} 
Correspondence and requests for materials should be addressed to Wai Kit Ng or Riccardo Sapienza.

\section*{Data availability} All data in support of this work is available in the manuscript or the supplementary materials. Further data and materials are available from the corresponding authors upon reasonable request.

\bibliography{references}


\onecolumngrid
\widetext
\appendix
\clearpage
\setstretch{2}
\begin{center}
  \textbf{\Large Supplementary information: \\
Mode visualisation and control of complex lasers using neural networks}\\[.2cm]
  Wai Kit Ng,$^{1}$ T. V. Raziman,$^{1,2}$ Dhruv Saxena,$^{1}$ Korneel Molkens,$^{3, 4, 5}$ Ivo Tanghe,$^{3, 4, 5}$ Zhenghe Xuan,$^{1}$ Pieter Geiregat,$^{3, 4}$ Dries Van Thourhout,$^{4, 5}$ Mauricio Barahona,$^{2}$ and Riccardo Sapienza$^{1}$ \\[.1cm]
  {\itshape ${}^1$Blackett Laboratory, Department of Physics,\\
  Imperial College London, London, UK\\
  ${}^2$Department of Mathematics, Imperial College London, London, UK\\
  ${}^3$Physics and Chemistry of Nanostructures (PCN),\\
  Ghent University, Krijgslaan 281-S3, B9000 Gent, Belgium\\
  ${}^4$Center for Nano- and Biophotonics,\\
  Ghent University, 9052 Ghent, Belgium\\
  ${}^5$Photonics Research Group, Ghent University - imec,\\
  Technologiepark-Zwijnaarde 126, 9052 Ghent, Belgium}\\[1cm]
\end{center}

\renewcommand{\appendixname}{}
\renewcommand{\thefigure}{S\arabic{figure}}
\renewcommand{\figurename}{Supplementary FIG.}
\setcounter{figure}{0}
\renewcommand{\theequation}{S\arabic{equation}}
\setcounter{equation}{0}
\renewcommand{\thesection}{Supplementary Note \arabic{section}}
\counterwithout{equation}{section}
\addtocounter{equation}{0}
\def\bibsection{\section*{Supplementary \refname}} 
\newpage

\fontsize{10}{11.5}\selectfont

\section{Characteristic of SiN-QDs microrings} \label{Sample_char}
The CdSe/CdS quantum dots (QDs) in a vertical SiN/QD/SiN stack structure supply gain to the microring lasers. The gain spectra of the CdSe/CdS QDs through transient absorption measurements are shown in Supplementary FIG. \ref{SUPPFIGURE_RingChars}(a). By structurally confining the emission, a standalone 10 $\rm \mu m$ microring lases efficiently under uniform illumination, with a few different lasing peaks corresponding to different transverse and longitudinal whispering gallery modes (WGMs) (Supplementary FIG. \ref{SUPPFIGURE_RingChars}(b)). 

The 10$\times$10 microring array presented in this work was designed and fabricated with slight variations in ring diameters and ring-to-ring gaps. The rings were designed to be $\mathrm{\sim 10~\mu m}$ in diameter with ring-to-ring gaps ranging from 0 -- 240 nm and a fixed $\mathrm{2 ~\mu m}$ width. The weak coupling from the disordered coupling strengths and detuned resonance frequencies across the array results in a complex lasing response (see FIG. \ref{FIGURE01}(b)). The statistical distributions of the ring diameters and gaps are shown in Supplementary FIG. \ref{SUPPFIGURE_RingChars}(c), (d).

\newpage
~\vfill
\begin{figure}[!ht]
    \centering
    \includegraphics[width= 0.85\textwidth]{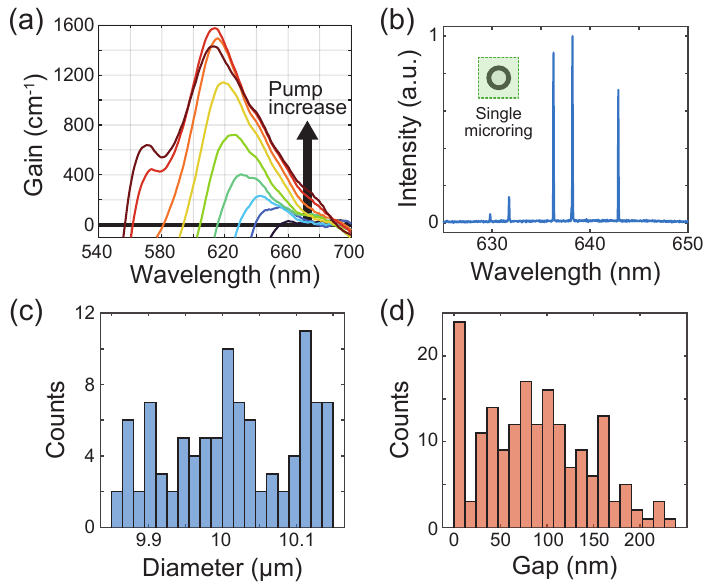}
    \caption{\textbf{Characterisations of the hybrid QD-SiN microring array.}
    (a) The gain spectrum of the CdSe/CdS quantum dots measured via transient absorption at 3 ps delay under different pump fluences \cite{bisschopImpactCoreShell2018}. (b) The emission spectrum of a standalone single microring laser with 10 $\mu$m diameter and 2 $\mu$m width. Multiple lasing modes are supported in the ring. (c, d) The statistical distributions of the (c) ring diameters and (d) ring-to-ring gaps in the 10$\times$10 disorderly coupled microring array.}\label{SUPPFIGURE_RingChars}
\end{figure}
\vfill
\newpage

\fontsize{10}{11.5}\selectfont

\section{Data preprocessing}\label{Data_preprocess}
Before feeding the data for the training of neural networks, the experimentally collected spectral data are pre-processed to simplify the classification task. Due to the photobleaching of the QDs gain under long illumination, the lasing spectra gradually offset across the data collection process. To correct for the shift of the emission spectrum, the spectrum of a reference excitation pattern (full illumination pattern) is collected for every 1000 data points during the data collection. The data are then wavelength-corrected based on the shift between the reference spectra. After that, the background noise is substrated from the wavelength-corrected spectra before we perform peak detection. Using the detected peak (mode) positions, the spectra data are converted into one-hot encoded mode data, which are vectors that only contain binary information on the existence of the modes (1 for lasing modes, 0 otherwise), to simplify the task in a multi-label classification problem.
\newpage

\fontsize{10}{11.5}\selectfont

\section{Architecture and training of the spectral prediction network (SN)}\label{SN_training}
The spectral prediction network (SN) is a densely connected artificial neural network (or MLP) with 2 hidden layers, as shown in Supplementary FIG. \ref{SUPPFIGURE_SN_detail}. Each pixel in an excitation pattern corresponds to a node in the input layer, and the input layer is connected to 2 hidden layers with 1024 and 64 nodes respectively. The processed information is propagated to the output layer with each node corresponding to a mode in the lasing spectra (one-hot encoded). For the 10$\times$10 microring array, the excitation beam is patterned in a 30$\times$26 grid that covers the entire array, with a total of 780 input pixels that can be switched on and off individually. By examining multiple lasing spectra, there are a total of 28 possible lasing modes identified.

A set of 7000 randomly generated Perlin noise patterns is illuminated on the sample individually to obtain their spectra under a fixed excitation fluence. The Perlin noise patterns have some degree of clustering of pixels, which is a better match with the gain distribution of photonic modes in general. This type of pattern can reduce the amount of sampling required for more efficient training. The lasing spectra collected from the pattern illuminations are first processed to extract the position and state (on or off) of the lasing peaks to simplify each lasing spectrum into a one-hot encoded ($28 \times 1$) mode vector. The 7000 pairs of mode vectors and patterns are randomly split into sets of 5950 (85\%) as a training dataset and 1050 (15\%) as a test dataset. The test dataset is untouched during the training process, and it is only used to evaluate the performance of the model. During the backpropagation training process, the connection weights between the nodes and the biases are adjusted based on the binary cross-entropy loss function calculated from the batches (batch size = 256) of the training dataset using the Adam optimisation algorithm (learning rate = 0.001). In each hidden layer, an L2 regulariser (with a decay rate of 0.001) is applied for ridge regression optimisation. This process is repeated until the network is well-trained such that no further improvements can be made by evaluating the test dataset. A total of 213 epochs have been trained, with a 93.68\% prediction accuracy for the test dataset. 
\newpage
~\vfill
\begin{figure}[!ht]
    \centering
    \includegraphics[width= 0.99\textwidth]{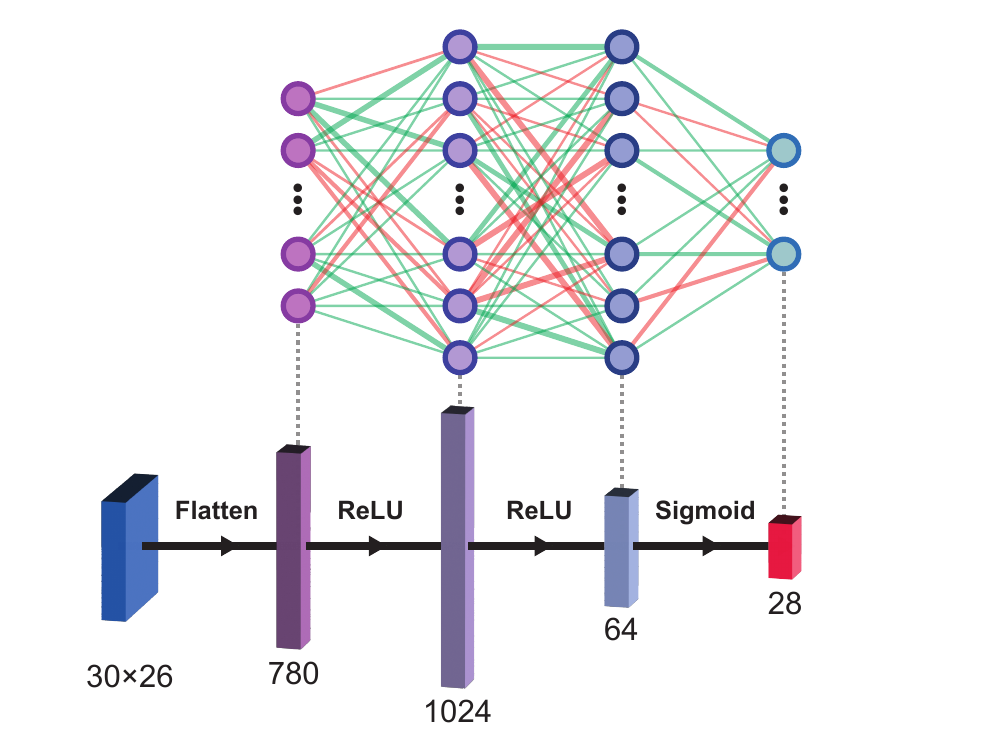}
    \caption{\textbf{Architecture of the spectral prediction network.} The 3-layer multi-label classification neural network model for mode visualisation. For the input and hidden layers (except the last hidden layer), a ReLU activation function is used. A sigmoid activation function is applied at the last hidden layer to produce a one-hot encoded mode spectral profile. The number of nodes in each layer is shown at the bottom}\label{SUPPFIGURE_SN_detail}
\end{figure}
\vfill
\newpage

\fontsize{10}{11.5}\selectfont

\section{Mode visualisation theory in multi-layer neural network}\label{MultiLayer_theory}

In a single-layer neural network, the neural network architecture is mathematically written as 
\begin{equation}\label{SingleNN_theory}
\mathbf{M} = \sigma(\mathbf{W} \cdot \mathbf{P} + \mathbf{B}), 
\end{equation}
where $\mathbf{P} = [p_{1}, ..., p_{I}]$ is the input layer and $\mathbf{M} = [m_{1}, ..., m_{M}]$ is the output layer of the neural network. The matrix $\mathbf{W}$ is the weight matrix describing the connection between $\mathbf{P}$ and $\mathbf{M}$, with the extra biases $\mathbf{B} = [b_{1}, ..., b_{M}]$. The mode visualisation theory for a single-layer neural network is described in the main text. In a 2-layer neural network, the complete architecture (without regularisations) can be written in the form of simultaneous equations, 
\begin{equation}\label{2LayerNN_sim_eq}
\left\{\begin{split}
\mathbf{M} & = \sigma_{2}(\mathbf{W''} \cdot \mathbf{H}+ \mathbf{B''})
\\ 
\mathbf{H} & = \sigma_{1}(\mathbf{W'} \cdot \mathbf{P} + \mathbf{B'})
\end{split}\right. ,
\end{equation}
where $\mathbf{W'}$ and $\mathbf{W''}$ are the weight matrices for the input-hidden and hidden-output layer connections. The activation functions $\sigma_{1}$ and $\sigma_{2}$ are the nonlinear functions applied to their corresponding layer. These simultaneous equations are equivalent to adding a layer $\mathbf{M}$ after the single-layer network with input $\mathbf{P}$ and output $\mathbf{H}$. Thus, the following 2-layer derivation can also be extended to N-layer neural networks by adding a layer to a (N-1)-layer neural network. 

In a 2-layer neural network, for a particular output node $m_{i}$, the sum of the connections from the input layer to $m_{i}$ is
\begin{equation}\label{2LayerNN_mi}
m_{i} = \sigma_{2}(\sum_{k} \mathbf{W''_{ik}} \sigma_{1}(\sum_{j} \mathbf{W'_{kj}}~p_{j} + b'_{k}) + b''_{i}).
\end{equation}
When considering the change of mode $m_{i}$ with respect to a spatial profile pixel $p_j$, the partial derivative $\frac{\partial m_{i}}{\partial p_{j}}$ can be expressed as
\begin{align}\label{2LayerNN_partialD_A}
\frac{\partial m_{i}}{\partial p_{j}} & = \frac{\partial[\sigma_{2}(\sum_{k} \mathbf{W''_{ik}} \sigma_{1}(\sum_{j} \mathbf{W'_{kj}}~p_{j} + b'_{k}) + b''_{i})]}{\partial p_{j}} \notag\\
& = \frac{\partial\sigma_{2}(u_{i})}{\partial u_{i}} \cdot \sum_{k} \mathbf{W''_{ik}} [\frac{\partial\sigma_{1}(v_{k})}{\partial v_{k}} \cdot \frac{\partial(\sum_{j} \mathbf{W'_{kj}}~p_{j} + b'_{k})}{\partial p_{j}}] \notag\\
& = \frac{\partial\sigma_{2}(u_{i})}{\partial u_{i}} \cdot \sum_{k} \mathbf{W''_{ik}} [\frac{\partial\sigma_{1}(v_{k})}{\partial v_{k}} \cdot \mathbf{W'_{kj}}],
\end{align}
in which the variables $u$ and $v$ are the dummy variables where $u_i = \sum_{k} \mathbf{W''_{ik}} \sigma_{1}(\sum_{j} \mathbf{W'_{kj}}~p_{j} + b'_{k}) + b''_{i}$ and $v_k = \sum_{j} \mathbf{W'_{kj}}~p_{j} + b'_{k}$. 

As in a single-layer neural network with different sets of values of $p_{i}$ ($i \neq j$), these partial derivative terms are different because the lasing behaviour of mode $m_{i}$ is also dependent on the gain/loss distribution in proximity due to mode competition. The partial derivative terms (with local activation functions), however, are still constant across each pixel $p_{j}$ for the same target mode $m_{i}$. To compare the contributions between pixels for the same mode, we assume the system is populated close to the threshold of mode $m_{i}$ such that the small increase in the pixel value $p_{j}$ can lead to the activation of $\sigma_{1}$ and $\sigma_{2}$ (in the hidden layer and output layer respectively) if the pixel $p_j$ is important to the mode. Both the partial derivative terms are then not critical factors as they will remain at a similar level for different pixels $p_{j}$. The contributions of the spatial profile pixel $p_j$ to mode $m_{i}$, therefore, can be estimated by
\begin{align}\label{2LayerNN_partialD_B}
\frac{\partial m_{i}}{\partial p_{j}} & \sim \sum_{k} \mathbf{W''_{ik}} \mathbf{W'_{kj}} = \mathbf{W_{ij}},
\end{align}
where $\mathbf{W} = \mathbf{W''}\cdot \mathbf{W'}$ is the equivalent weight matrix connecting input and output layers. Hence, the spatial gain profile of mode $m_{i}$ can be estimated by the $i^{th}$ row of the equivalent weight matrix $\mathbf{W}$. 

\newpage

\fontsize{10}{11.5}\selectfont
\section{Training of the tandem neural network}\label{TNN_training}
The tandem neural network (TNN) model is initialised with an untrained control network (CN) and a well-trained spectral prediction network (SN) joined in series. The SN is pre-trained with the 5950 sets of experimental data in the way described in the \ref{SN_training}. During the TNN training, only the weights and biases in CN are updated, and the SN is treated as an untrainable network. The same training dataset used for the SN training is also used to train the TNN. When optimising the network to achieve good predictions for both the existence of spectral mode and the excitation pattern, the performance and the loss of the model are evaluated in the whole TNN. To achieve a realistic excitation pattern prediction from the target mode, the training takes into consideration both the spectral and spatial accuracies with an assigned complex loss function, 
\begin{equation}\label{eq_loss}
\begin{split}
\textup{loss} = & -\frac{1}{N}\sum_{n=1}^{N} \sum_{m=1}^{M} \left[ S_{nm} \log S'_{nm} +  (1 - S_{nm}) \log(1 - S'_{nm}) \right] \\
& + \alpha \cdot \frac{1}{N} \sum_{n=1}^{N} \sum_{i=1}^{I} (P_{ni} - P'_{ni})^{2} ,
\end{split}
\end{equation}
where $N = 5950$ is the size of the training dataset, $M = 28$ is the total number of modes, and $I = 780$ is the number of pixels in the patterns. $S_{nm}$ ($S'_{nm}$) and $P_{ni}$ ($P'_{ni}$) are the probabilities of the spectral mode and the spatial pattern pixel in the ground truth (prediction), respectively. The first term in this complex loss function is the binary cross-entropy function of the spectra, which is the same as the loss function in the SN training for mode visualisation. The second term is the mean-squared error function of the predicted excitation patterns. The constant $\alpha$ is a hyperparameter to adjust the model to optimise more towards either pattern similarities or mode accuracy. Here, the value of $\alpha = 0.25$ was chosen to obtain a good mode prediction accuracy while keeping the excitation pattern physical.
\newpage

\end{document}